\documentclass[submission, Phys]{SciPost}

\hypersetup{colorlinks,linktocpage,citecolor=blue,linkcolor=red,urlcolor=blue}
\usepackage[capitalize]{cleveref}
\crefformat{section}{Sec.~#2#1#3}

\usepackage{subcaption}
\usepackage{amsmath, amsfonts}
\usepackage{mcite}
\usepackage{mathrsfs}

\graphicspath{{figures/}}

\numberwithin{equation}{section}
\usepackage{amsmath,amssymb}
\usepackage{tipa}
\usepackage{upgreek}
\usepackage{comment}
\usepackage{graphicx}
\usepackage{braket}
\usepackage{enumitem}
\usepackage{mathbbol}
\usepackage{color}

\newcommand\x{\mathbf{x}}
\renewcommand\d{\partial}
\let\vec\mathbf

\def\d{\partial}

\def\cR{\mathcal{R}}

\def\cS{\mathcal{S}}

\def\cK{\mathcal{K}}
\def\cR{\mathcal{R}}

\def\>{\rangle}
\def\<{\langle}

\renewcommand\d{\partial}

\newcommand\y{\mathbf{y}}

\newcommand\ml{\mathbf{l}}
\newcommand\mQ{\mathbf{Q}}
\newcommand\ms{\mathbf{s}}
\newcommand\mk{\mathbf{k}}
\newcommand\mx{\mathbf{x}}
\newcommand\my{\mathbf{y}}
\newcommand\mq{\mathbf{q}}

\newcommand\mK{\mathbf{K}}
\newcommand\mR{\mathbf{R}}

\newcommand{\be}{\begin{equation}}
\newcommand{\ee}{\end{equation}}
\newcommand{\bea}{\begin{eqnarray}}
\newcommand{\eea}{\end{eqnarray}}

\begin{document}
	\begin{center}{\Large \textbf{
				Landau levels in curved space realized in strained graphene
	}}\end{center}
	
	\begin{center}
		Glenn Wagner  \textsuperscript{1},
		Fernando de Juan\textsuperscript{2,3},
		Dung X. Nguyen\textsuperscript{4,*}
	\end{center}
	
	\begin{center}
		{\bf 1} Rudolf Peierls Centre for Theoretical Physics, Parks Road, Oxford, OX1 3PU, UK
		\\
		{\bf 2} Donostia International Physics Center, Paseo Manuel de Lardizabal,
		4, 20018 San Sebastian, Spain
		\\
		{\bf 3} IKERBASQUE, Basque Foundation for Science, Maria Diaz de Haro 3, 48013 Bilbao, Spain
		\\
			{\bf 4} Brown Theoretical Physics Center and Department of Physics,
			Brown University, 182 Hope Street, Providence, RI 02912, USA\\
		* \href{mailto:dungmuop@gmail.com}{dungmuop@gmail.com}
	\end{center}
	
	\begin{center}
		\today
	\end{center}

	{\bf
	\section*{Abstract}
	The quantum Hall effect in curved space has been the subject of many theoretical investigations in the past, but devising a physical system to observe this effect is hard. Many works have indicated that electronic excitations in strained graphene realize Dirac fermions in curved space in the presence of a background pseudo-gauge field, providing an ideal playground for this. However, the absence of a direct matching between a numerical, strained tight-binding calculation of an observable and the corresponding curved space prediction has hindered realistic predictions. In this work, we provide this matching by deriving the low-energy Hamiltonian from the tight-binding model analytically to second order in the strain and mapping it to the curved-space Dirac equation. Using a strain profile that produces a constant pseudo-magnetic field and a constant curvature, we compute the Landau level spectrum with real-space numerical tight-binding calculations and find excellent agreement with the prediction of the quantum Hall effect in curved space. We conclude discussing experimental schemes for measuring this effect.
}
	
	\vspace{10pt}
	\noindent\rule{\textwidth}{1pt}
	\tableofcontents\thispagestyle{fancy}
	\noindent\rule{\textwidth}{1pt}
	\vspace{10pt}
	
	
\section{Introduction}

When two dimensional electrons are subject to a magnetic field, one of the best known quantum effects in solids is displayed. Electronic wavefunctions are confined to cyclotron orbits with discrete energy levels known as Landau levels, and the quantum Hall effect is famously observed \cite{cage2012quantum}. While electrons with parabolic dispersion as found in semiconductors display equally spaced Landau levels, the advent of graphene brought to prominence the fact that massless Dirac fermions display a different type of Landau quantization with $E_n = v_F \sqrt{2nB}$, where $v_F$ is the Fermi velocity, which has then been used as proof of the existence of Dirac quasiparticles in many systems beyond graphene ever since \cite{vafek2014dirac}. 

Interestingly, a new type of Landau quantization was predicted long ago for Dirac fermions living in a background curved space with constant Riemannian curvature $\cK$ ~\cite{Pnueli:1994}. While the dynamics of Dirac electrons in curved spaces is complex, the presence of constant curvature and magnetic fields leads to a remarkably simple Landau level spectrum of the form
\begin{equation}
	E_n(B,\cK)=
	v_F \textrm{sgn}(n)\sqrt{2|nB|+n^2\cK},
	\label{eq:energy_levels}
\end{equation}  
valid for $\frac{|B|}{\cK}>\frac{1}{2}$ if $\cK>0$, and for $\frac{|B|}{|\cK|}>(|n|+\frac{1}{2})$ if $\cK<0$. This prediction is of great interest because it allows one to probe the effects of Riemann curvature on Dirac fermions by measuring their spectrum in magnetic field, but it has remained untested because of the difficulty in realizing a physical system with Dirac fermions in constant $B$ and $\cK$.

Among the possible systems available, strained graphene is actually one of the better suited physical candidates to observe this type of physics \cite{Fullerene,Nanoribbons,Nanoribbons2,defect1,defect2}. In the presence of strain, the low energy Dirac electrons in graphene behave as though they feel an effective magnetic field and move in a curved space-time \cite{Juan:2012,Juan:2013,OLIVALEYVA20152645,Zubkov:2015,Oliva_Leyva_2017,Zubkov2,SpinConnection2,SpinConnection3,SpinConnection4,Straintronics_Review,Castro-Villareal17}. Indeed, for strains where the magnetic field is constant, a Landau level spectrum should develop, which has been shown in tight-binding calculations \cite{Guinea2009,Settnes} and in experiment \cite{Levy544,YEH20111649,Lu2012,Li2015}. Since strain should serve both to provide the magnetic field and the curvature, without the need of an external magnetic field, this makes this system an ideal playground to study Dirac fermions in curved space. We note that the surface states of topological insulators, which also realize Dirac fermions, have often been discussed in this context as well, since the surface can take any shape and is in general curved \cite{DungHai09,Zhang10,Dahlhaus10,Parente11,Takane13,deJuanWires19,Kozlovsky20}. The magnetic field, however, has to be supplied externally and its spatial dependence is fixed. 

As promising as the graphene platform might be, elaborating on its microscopic model illustrates an unsolved challenge. Electrons hopping on the graphene honeycomb lattice can be described by a tight-binding (TB) model \cite{Review}, and its continuum limit provides the Dirac fermion action naturally \cite{fradkin2013field}. When strain is applied, the local hopping strength varies across the lattice, and these leads to two corrections to the Dirac action\cite{Juan:2012,Juan:2013,OLIVALEYVA20152645,Zubkov:2015,Oliva_Leyva_2017,symmetry1,symmetry2,symmetry3}: An emergent space-dependent gauge field (and thus a magnetic field) and a space dependent Fermi velocity tensor, which can equally be interpreted as a the vielbein coupling fermions to a curved space. Given these ingredients, it appears that an exact mapping between the tight-binding model of strained graphene and the corresponding continuum field theory must exist, which would allow to find the strain profile that realizes any desired curved space. Since certain aspects of this mapping are quite subtle, it has remained a challenge to produce a real space lattice calculation whose spectrum exactly matches a field theory prediction in curved space. The subtleties include the momentum origin of the low-energy expansion, the different orders of expansion in strain, how to deal with the volume form when comparing with tight binding and whether a spin-connection term should appear in the description.

In this work, we derive the mentioned mapping at the lattice level to all orders in strain, and we take the simple field theory prediction for Landau levels in constant curvature, Eq. \ref{eq:energy_levels} as a benchmark to test it. Using the mapping to choose an appropriate strain profile, and expanding to second order in strain, we show that the spectrum of a real space lattice calculation with modulated hopping displays the expected series of Landau levels, with their positions matching accurately the prediction from the field theory in curved space, Eq. \ref{eq:energy_levels}. We emphasize we realize effective curvature coming purely from the in-plane strain, in contrast to previous work considering physically curving the graphene sheet \cite{Curved}. Our numerical work therefore provides support for our precise mapping, and shows the strained honeycomb lattice is indeed an ideal system to realize predictions for Dirac fermions in curved space. 

Besides the potential experimental realization in solid state graphene, we propose that such a strain profile can be engineered---and the effects of curvature seen---in graphene analogues such as photonic or sonic lattices, where the level of control is much higher than in conventional graphene. In these systems it is possible to modify the position of each lattice site individually. It would be exciting indeed to see the effects of curvature on the quantum Hall effect in these experimental set-ups. Synthetic LL for photons have already been seen in optical resonators \cite{Gromov1,Gromov2}.	

The structure of our paper is the following: In Sec. \ref{sec:FT} we introduce the field theory formalism of a Dirac fermion in curved space. Sec. \ref{sec:TB} shows the explicit map between the continuum Hamiltonian derived from the TB model and the field theory from Sec. \ref{sec:FT}. We then focus on a particular strain profile for Sec. \ref{sec:2nd_order} and use the expressions from Sec. \ref{sec:TB} to evaluate the magnetic field and curvature to second order in strain for this particular profile. We perform numerics on this TB model in Sec. \ref{sec:num} and see explicitly the effects of curvature in the results. In Sec. \ref{sec:comparison} we present a comparison of our results with other works.

\section{$2+1 D$ Dirac Fermion in static curved space}
\label{sec:FT}
In this section, we will review the general form of the Hamiltonian of a $2+1D$ Dirac fermion in static curved space following the textbook by Bertlmann \cite{Bertlmann:2000}.  It can be seen that we do not have any term corresponding to the spin connection explicitly.
\subsection{Vielbein formalism}

Let us first introduce some notation that will be useful later on. In particular, we use the vielbein formalism of General Relativity \cite{Wald:GR,Weinberg}, which exploits the fact that there is always a coordinate transformation to a locally flat frame. 
In our work we choose to focus on the case where the atoms in the graphene sheet only experience in-plane displacements and hence we end up with a $2+1D$ problem. The metric of static curved space is given by 
\begin{equation}
	\label{eq:metricsc}
	g_{\mu \nu}(\mx)=\begin{pmatrix} 
		1 & 0 \\
		0 &   -g_{ij}(\mx)\\
	\end{pmatrix},
\end{equation}
where $i,j,k=1,2$ represent the space indices.  The flat space-time metric is $
\eta_{\alpha \beta}=\textrm{diag}(1,-1,-1)$. We will use $\alpha,\beta,\gamma,\cdots$ for local frame indices and $\mu,\nu,\lambda, \cdots$ for coordinate indices. We will also use $a,b,c,\cdots=1,2$ for space indices of the local frame and $i,j,k,\cdots=1,2$ for space indices of the coordinates. The vielbein is given by the definition $
g_{\mu\nu}=e^\alpha_\mu e^\beta_\nu \eta_{\alpha\beta}$. We raise and lower the local frame indices by $\eta_{\alpha\beta}$ and $\eta^{\alpha \beta}$, we raise and lower the space coordinate indices by $g_{\mu\nu}$ and $g^{\mu\nu}$. We also define the inverse vielbein $e^\mu_\alpha$ as $
e^\alpha_\mu e^\mu_\beta=\delta^{\alpha}_\beta$. For static curved space, we have \begin{equation}
	\label{eq:stacurv}
	e^0_0=1, e_0^i=e_a^0=0 ,
\end{equation}
and all the components of the vielbein are time-independent. The spin connection is \footnote{We consider a torsionless spin connection and smooth valued vielbeins. The field theory in curved space with torsion is required in the presence of lattice defects such as dislocations and disclinations, see for example \cite{Hughes:Torsion,Carlos:Torsion,SpinConnection1}.}
\begin{equation}
	\label{eq:spinconect}
	\omega_{\mu\beta}^{\alpha} =-e^\nu_\beta(\partial_\mu e^\alpha_\nu-\Gamma^\sigma_{\mu\nu}e^\alpha_\sigma).
\end{equation}
where $\Gamma^\mu_{\lambda\sigma}$ is the Christoffel symbol. With the tetrad postulate \cite{Wald:GR}
\begin{equation}
	\label{eq:tetrad}
	\nabla_\mu e^\alpha_\nu=\partial_\mu e^\alpha_\nu-\Gamma^\sigma_{\mu\nu}e^\alpha_\sigma+\omega_{\mu \beta}^{\alpha}e^\beta_\nu=0,
\end{equation}
we find that the Christoffel symbol in terms of the vielbein is $\Gamma^\mu_{\lambda\sigma}=\d_\sigma (e_\lambda^\alpha) e^{\mu}_\alpha.$
We also use the explicit notation of gamma matrices with local frame index $\upgamma^\alpha$ in $2+1D$ as
\begin{equation}
	\upgamma^0=\sigma^3, \qquad \upgamma^a=\sigma^3 \sigma^a \qquad (a=1,2),
\end{equation}
where $\sigma^a$ are the Pauli matrices. We have the anti-commutation relation $
\left\lbrace \upgamma^a,\upgamma^b\right\rbrace =2 \eta^{ab}I$.	The space-dependent gamma matrices with space-time indices are given by 
\begin{equation}
	\label{eq:defgamma}
	\gamma^{\mu}(\mx)=e_\alpha^\mu(\mx)\upgamma^\alpha
\end{equation}
with the anti-commutation relation $
\left\lbrace \gamma^\mu(\mx),\gamma^\nu(\mx)\right\rbrace =2 g^{\mu\nu}(\mx)I$.

\subsection{$2+1D$ Dirac fermion in static curved space}
\label{sec:Dirac_curv}
The action of a spin-$\frac{1}{2}$ Dirac fermion in $2+1D$ curved space-time is most often written as \cite{Wald:GR,Weinberg,frankel:GT,Bertlmann:2000} 
\begin{equation}
	\label{eq:LagNonHerm}
	\cS=i\int d^3 x \sqrt{|g|}\bar{\Psi} \gamma^\mu(\partial_\mu-iA_\mu-\frac{i}{2}\omega_\mu^{\alpha \beta}\sigma_{\alpha\beta})\Psi
\end{equation}
where we used $
\sigma_{\alpha\beta}=\frac{i}{4}\left[\upgamma_\alpha,\upgamma_\beta\right],$	and $g=\det (g_{\mu\nu})$, and $\bar{\Psi}=\Psi^\dagger \upgamma^0$. $\omega_\mu^{\alpha \beta}$ is the spin connection. This action in Eq. \eqref{eq:LagNonHerm}, as written, is formally not real, which implies that the corresponding Hamiltonian is not Hermitian. A properly real action is rather obtained as \cite{Bertlmann:2000}
\begin{equation}
	\label{eq:Lagcur0Herm}
	\cS'=\frac{1}{2}\left(\cS+\cS^\dagger\right).
\end{equation}
However, $\cS$ and $\cS'$ only differ by a total derivative (see Appendix \ref{sec:DetailDirac} for a proof) and are hence equivalent physically in an infinite system, and because of this $\cS$ or $\cS'$ are used interchangeably in the literature. This also implies that $\cS$ is in fact also real in an infinite system, upon discarding a vanishing boundary term. In the presence of a boundary, it is of course $\cS'$ that should be used, or if $\cS$ is used it should be supplemented with a boundary condition that makes the whole action Hermitian \footnote{In fact the natural choice of boundary conditions is the one that makes the different surface terms vanish, see Appendix \ref{sec:DetailDirac}1}.In this work we will just use $\cS'$ for simplicity. As a real action leading to a Hermitian Hamiltonian, this is also the natural choice to compare with a lattice tight binding Hamiltonian.

%
%
In a static curved space \eqref{eq:metricsc}, the Hermitian Hamiltonian corresponding to \eqref{eq:Lagcur0Herm} is
\begin{equation}
	\label{eq:Hamcur0}
	H=-i v_F\int d^2 \mx \sqrt{\hat{g}}\left[\Psi^\dagger e^i_a \sigma^a\left(\overleftrightarrow{\partial}_i-iA_i\right)\Psi\right],
\end{equation}
where $\overleftrightarrow{\partial}_\mu=\frac{1}{2}\left(\overrightarrow{\d}_\mu-\overleftarrow{\d}_\mu\right)$ only acts on fermion fields. One observes that the spin connection term in \eqref{eq:Hamcur0} has cancelled for a static curved metric \eqref{eq:metricsc}, which is a well known result for $2+1D$ Dirac fermions \cite{Bertlmann:2000}. However, the equation of motion derived from the Hermitian Hamiltonian \eqref{eq:Hamcur0} will have an additional term which includes the derivative of the vielbein. This term originates from the spin connection. See Appendix \ref{sec:DetailDirac} for more details. Aiming at the comparison with graphene, we have inserted the Fermi velocity $v_F$ as appropriate for graphene, and $\hat{g}(\mx)=\det (g_{ij}(\mx))$. Furthermore, we have set the scalar potential to zero as is appropriate for the type of strain that we consider in this work. We leave the detailed derivation of equation \eqref{eq:Hamcur0} to Appendix \ref{sec:DetailDirac}.

The Hermitian Hamiltonian in Eq. \eqref{eq:Hamcur0} is not yet ready to be compared with the tight binding Hamiltonian. The reason is that the inner product of the wave function in static curved space is given by 
\begin{equation}
	\label{eq:inprod0}
	\<\Phi|\Psi\>=\int d^2 x \sqrt{\hat{g}} \Phi^\dagger \Psi.
\end{equation}
However, to make a comparison with the tight binding model we need to define a new field
\begin{equation}
	\label{eq:rename}
	\tilde{\Psi}=\hat{g}^{1/4}\Psi,
\end{equation}
with the corresponding wave function's inner product 
\begin{equation}
	\label{eq:inprod1}
	\<\tilde{\Phi}|\tilde\Psi\>=\int d^2 x \tilde{\Phi}^\dagger \tilde{\Psi},
\end{equation}
since this inner product is the one that will be inherited from the tight-binding model. The conjugate anti-commutation relation of Dirac fermion in curved space derived from the action $\mathcal{S}'$ is 
\begin{equation}
    \lbrace \Psi^\dagger(\mathbf{x}),\Psi(\mathbf{y}) \rbrace=\frac{1}{\sqrt{\hat{g}(\mathbf{x})}}\delta(\mathbf{x}-\mathbf{y}).
\end{equation}
After the scaling, we arrive at a new anti-commutation relation 
\begin{equation}
    \lbrace \tilde{\Psi}^\dagger(\mathbf{x}),\tilde{\Psi}(\mathbf{y}) \rbrace=\delta(\mathbf{x}-\mathbf{y})
\end{equation}
which is identical to the anti-commutation relation of the electron field in graphene. Therefore one should map the electron field in the tight binding Hamiltonian to the rescaled Dirac field $\tilde{\Psi}$ instead of $\Psi$. 
Note a similar rescaling has been used in Ref. \cite{Kozlovsky20}. The Hamiltonian in static curved space has the new form
\begin{equation}
	\label{eq:Hamcur}
	\tilde{H}=-iv_F \int d^2 x \tilde{\Psi}^\dagger e^i_a\sigma^a\left(\overleftrightarrow{\partial}_i-i A_i\right)\tilde{\Psi}.
\end{equation}	
In the subsequent sections, we will derive the same Hamiltonian \eqref{eq:Hamcur} from strained graphene. Naturally, the inner product of the electron's wave function in graphene takes the form \eqref{eq:inprod1}.  

So in conclusion, in order to compare to the tight-binding Hamiltonian, one has to use the Hermitian Hamiltonian \eqref{eq:Hamcur} instead of the Hamiltonian derived directly from \eqref{eq:LagNonHerm}. There is no explicit spin connection term in \eqref{eq:Hamcur}.

\section{Deriving the Dirac Hamiltonian from the tight-binding model}
\label{sec:TB}
\subsection{Tight-binding Hamiltonian}
Under applied strain, the tunneling parameter in the tight binding (TB) model of graphene depends on the coordinates. We write the TB Hamiltonian in the following form
\begin{equation}
	\label{eq:TB0}
	H^{TB}=\sum_{n, \mathbf{R}_i}t_n(\mathbf{R}_i)\left(\psi_A^{\dagger}(\mathbf{R}_i)\psi_B(\mathbf{R}_i+\mathbf{l}_n)+h.c.\right),
\end{equation}
where $t_n(\vec x)$ is the space dependent tunneling parameter, $\psi_A^\dagger(\mathbf{R}_i)$ is the operator that creates an electron at position $\mathbf{R}_i$ of sub-lattice $A$ and $\psi_B(\mathbf{R}_i+\mathbf{l}_n)$ is the operator that annihilates an electron at position $\mathbf{R}_i+\mathbf{l}_n$ of sub-lattice $B$. $\mathbf{l}_n$ are vectors from an $A$ site to its nearest neighbours, $\mathbf{l}_1=a(\frac{\sqrt{3}}{2},\frac{1}{2})$,  $\mathbf{l}_2=a(-\frac{\sqrt{3}}{2},\frac{1}{2})$ and $\mathbf{l}_3=a(0,-1)$. 

We Fourier transform the Hamiltonian \eqref{eq:TB0} and expand for momenta close to the Dirac point  $\mK=(\frac{4\pi}{3\sqrt{3}a},0)$. We note that there has been some controversy in the literature about whether to expand the Hamiltonian about the unshifted Dirac point  $\mK$ (as we do here) or about a shifted Dirac point, defined as the point in $k$-space where the Hamiltonian vanishes \cite{Zubkov:2015}. However, following through the expansion around the shifted $\mK$ point in the formalism, in Appendix \ref{sec:A_K_pt} one encounters divergences when attempting to Fourier transform back to real space, showing that this expansion is inconsistent. See Appendix \ref{sec:A_K_pt} for details.

We expand to linear order in the momenta while keeping the full hoppings $t_n(\vec x)$ which might have an arbitrary strain dependence for now. Going back to position space, we can rewrite the Hamiltonian as 
\begin{align}
	\label{eq:HamCurK}
	H^{TB}=&\int d^2 \x v_F\psi^\dagger(\x)\left(-i\sigma_i \tilde{v}^{ij}(\x)\overleftrightarrow{\d}_j -\sigma^i A^{s}_i (\x) \right)\psi(\x)
\end{align} 
with $v_F=\frac{3 t_0}{2 a}$ and the space dependent Fermi velocity is
\begin{align}
	\label{eq:vb}
	\tilde{v}^{ij}(\x)=\sum_{n}\frac{2}{3t_0 a^2}l_n^il_n^j\left(t_n(\x)-\frac{1}{2}l^k_n \d_k t_n(\x)\right),
\end{align}
and the artificial gauge field due to the strain is given by
\begin{equation}
	\label{eq:At}
	A^{s}_i(\x)=\sum_{n}\frac{2}{3 t_0 a^2}\epsilon_{ij}l^j_n\left(t_n(\x)-\frac{1}{2}l^k_n \d_k t_n(\x)\right)
\end{equation}
and $\psi (\x)$ is a Dirac spinor with the spinor index corresponding to the sub-lattice index 
\begin{equation}
	\label{eq:psiK}\psi(\mx)=e^{i \mK\cdot \x}\begin{pmatrix}
		\psi_A(\mx)\\ \psi_B(\mx) \end{pmatrix}.
\end{equation}
The result \eqref{eq:HamCurK} together with \eqref{eq:vb} and \eqref{eq:At} are central results of this work. The technical details of this calculation are left to the Appendix \ref{sec:TBApx}, where we also discuss how the last terms of \eqref{eq:vb} and \eqref{eq:At} were missed in the previous works \cite{Juan:2012,Juan:2013} and the implications of this. Furthermore, in this appendix we show that there is a symmetry operation relating the low-energy theory around $\vec K$ and around the other Dirac point $\vec K'=-\vec K$, such that the spectra are the same.
\subsection{Mapping to Dirac fermion in static curved space}
\label{sec:map}
We will now relate the TB Hamiltonian to the Dirac Hamiltonian in curved space derived in section \ref{sec:Dirac_curv}. Since the local velocity $\tilde{v}_{ij}$ is symmetric by construction, we can rename it as follows
\begin{equation}
	\tilde{v}_{aj}\rightarrow \tilde{v}_{a}^j \qquad (a=1,2; \ i=1,2).
\end{equation}
We can rewrite the tight binding Hamiltonian as \footnote{By renaming $i$ to $a$.}
\begin{equation}
	\label{eq:HamTBmap}
	H^{TB}=v_F\int d^2 \mx \psi^{\dagger}\left(-i \tilde{v}_{a}^i(\mx)\sigma^a \overleftrightarrow{\partial_i}  -\sigma^a A^s_a(\x)\right)\psi.
\end{equation}
Comparing the TB Hamiltonian 	\eqref{eq:HamTBmap} and the Dirac Hamiltonian in static curved space \eqref{eq:Hamcur} we obtain the explicit map
\begin{align}
	\label{eq:map}
	e_a^i(\x)&=\tilde{v}_{a i}(\x),\\  \label{eq:map1} A_i(\x) &=e^a_i(\x)A^s_a(\x), \\ \tilde{\Psi}(\x)&=\psi(\x),	
	\label{eq:map2}
\end{align}
where $e^a_i(\x)$ is the inverse vielbein such that  $e^a_i(\x) e^i_b(\mx)=\delta^a_b$. Note that $\tilde{\Psi}(\x)$ and $\psi(\x)$ share the same definition of inner product (without $\sqrt{\hat{g}}$). This mapping confirms that the low-energy graphene Hamiltonian in the presence of space-dependent tunneling induced by strain can always be mapped to a field theory of a Dirac fermion in static curved space and in magnetic field.  

Explicitly, the values of the curvature and magnetic field can be obtained from the local tunneling $t_n(\vec x)$ as follows. From Eq. \eqref{eq:vb} and the inverse vielbein $e^a_i(\mx)$, one can derive the spatial component of the metric $g_{ij}(\mx)=e^a_{i}(\mx)e^a_j(\mx)$. The Gaussian curvature in two dimensions can be calculated using the formula \cite{frankel:GT}
\begin{equation}
	\cK(\x)=\frac{\cR_{1212}(\mx)}{\hat{g}(\mx)},
	\label{eq:curv_2d}
\end{equation} 
where the Riemann curvature tensor is given by
\begin{equation}
	\cR_{ijkl}=g_{lq}\left(\d_i \Gamma^{q}_{jk}-\d_j \Gamma^{q}_{ik}+\Gamma_{jk}^p\Gamma^{q}_{ip}-\Gamma_{ik}^p\Gamma^{q}_{jp}\right).
\end{equation}
The magnetic field in static curved space can be obtained from the vector potential in Eq. (\ref{eq:At}) as
\begin{equation}
	B(\x)=\frac{\d_1 A_2(\x)-\d_2 A_1(\x)}{\sqrt{\hat{g}(\x)}}.
	\label{eq:B_field}
\end{equation}

In this paper, we only consider the strain profiles with the condition $\partial_i u^i=0$, in which $u^i$ is the strain field defined by the displacement $x^i \to x^i+u^i(\mathbf{x})$. With these strain profiles, there is no strain induced scalar potential. However, with a general strain, there would be an extra term $-A_0 \psi^\dagger \psi$ contributing to the tight binding Hamiltonian where the scalar potential is given by \cite{Vozmediano:2010PhysRep} 
\begin{equation}
\label{eq:scalar}
A_0=g \partial_i u^i,	
\end{equation}
where $g$ is a constant that can be determined by numerical calculation \cite{Vozmediano:2010PhysRep}. Consequently, the tight binding Hamiltonian will be mapped to a more general Dirac Hamiltonian in static curved space with both vector potential and scalar potential \eqref{eq:HamGen} with 
gauge potential $A_\mu$ and vielbein $e^i_a$ given by \eqref{eq:scalar}, \eqref{eq:map1} and \eqref{eq:map}.

We are now in a position to devise a lattice deformation to produce constant magnetic field and curvature. To do so, we next consider expanding $t_n(\vec x)$ to second order in strain.\\

\section{Constant curvature and magnetic field from strain}	
\label{sec:2nd_order}
We strain the lattice such that the positions of the lattice sites are displaced as 
\begin{equation}
	\vec{x}\to\vec{x}+\vec u(\vec x).
\end{equation}
\begin{figure}
\centering
	\includegraphics[width=0.5\columnwidth]{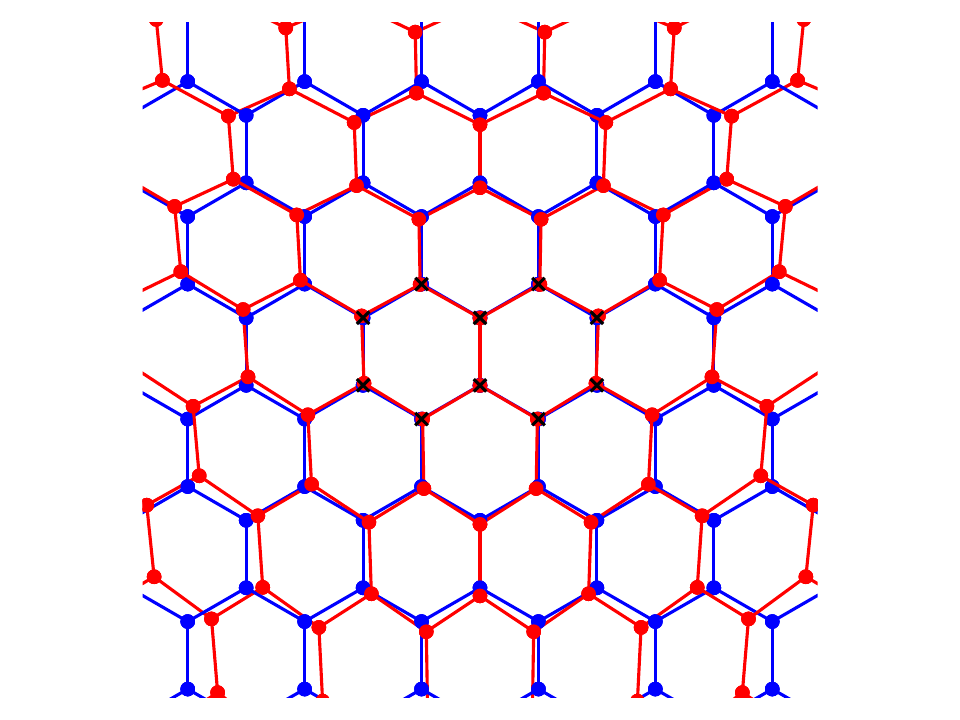}
	\caption{Sketch of the central region of the lattice. The undistorted honeycomb lattice is shown in blue, the strained lattice (with strain profile \eqref{eq:strain_profile}) is shown in red. The 10 central sites that are used to calculate the LDOS are marked by a black cross. }
	\label{fig:lattice}
\end{figure}
Let us propose the following strain profile (shown in Fig.~\ref{fig:lattice}), which produces a constant magnetic field at first order in the strain
\begin{equation}
	\label{eq:strain_profile}
	\vec u=\frac{u_B}{L}\begin{pmatrix}
		2xy\\
		x^2-y^2
	\end{pmatrix}
\end{equation}
where $u_B$ is a dimensionless number and $L$ is the total size of the lattice. For our approximations to hold, we want the change in the distance of neighbouring atoms due to strain to be less than the lattice spacing $a$. This implies 
\begin{equation}
	a\nabla \vec{u}\ll a
\end{equation}
and if we want this to hold up to the edge of the lattice we therefore require $u_B\ll1$. The hopping usually depends on the distance between nearest-neighbour atoms as 
\begin{equation}
	t_n(\vec x)=t_0e^{-\beta(|\vec l_n+\vec u(\vec x+\vec l_n)-\vec u(\vec x)|-a)/a},
	\label{eq:tn_exp}
\end{equation}
where $\beta\approx3$ for real graphene \cite{Ribeiro_2009}. For the photonic lattice, this form is also valid and $\beta$ can be tuned. For the purpose of a cleaner numerical match with the field theory, we will rather use a hopping profile that is exactly linear in the displacement with vanishing higher order terms
\begin{equation}
	t_n(\vec x)=t_0(1-\frac{\beta}{a^2} (\vec u(\vec x+\vec l_n)-\vec u(\vec x))\cdot \mathbf{l}_n)
	\label{eq:tn_lin}
\end{equation}
while still carrying out the rest of the calculation to second order in strain. The expression \eqref{eq:tn_lin} for the tunneling results in cleaner LL compared to using the full exponential profile \eqref{eq:tn_exp}. So although this form of the tunneling is not exactly the form that is relevant to experiments, we use it to check that our TB numerics reproduce the field theory predictions. It has already been noted in \cite{Linear1,Linear2} that this strain profile results in cleaner LL.

The detailed derivation of the space-dependent Fermi velocity $\tilde v^{ij}(\vec x)$ and the gauge field $A_a^s(\vec x)$ for the tunneling \eqref{eq:tn_lin} is found in Appendix \ref{sec:App_2nd_order_lin}: the final expressions are \eqref{eq:App_vF_lin} and \eqref{eq:App_gauge_field_lin} respectively. We apply the map \eqref{eq:map}-\eqref{eq:map2} and then use the formulae for the curvature \eqref{eq:curv_2d} and the magnetic field \eqref{eq:B_field} from the previous section. For this strain profile, we find curvature
\begin{equation}
	\label{eq:result_for_K}
	\cK=-\frac{4}{a^2}\bigg(\beta\frac{au_B}{L}\bigg)^2
\end{equation}
and magnetic field	
\begin{equation}
	\label{eq:result_for_B}
	B=\frac{4}{a^2}\beta\frac{au_B}{L}
\end{equation}
We see that we have both a constant curvature and constant magnetic field, as desired. 

We can also perform the same calculation for the exponential tunneling \eqref{eq:tn_exp}. We need to expand to second order in the strain, since we want to investigate the effect of curvature on the Landau levels, which appears at that order for the given profile. The calculation is performed in Appendix \ref{sec:App_2nd_order}. The expressions for the curvature and magnetic field in that case are given by equations \eqref{eq:K_exp} and \eqref{eq:B_exp} respectively. We emphasize that our approach can be used to compute $\cK$ and $B$ to arbitrary order in the strain. Beyond the second order we observe that the curvature and magnetic field are no longer constant in space. 

We note that it is possible to have non-zero curvature at first order in the strain if $\nabla\cdot\vec u\neq0$, however in this case the distance between neighbouring sites will significantly differ from $a$ far away from the centre of the lattice. In addition this leads to the scalar potential we neglected. For the numerical calculation this results in blurred out LL.

\section{Numerical calculation} 
\label{sec:num}
By exactly diagonalizing the Hamiltonian, we can determine the spectrum and compare to the field theory calculation. We use a lattice with around 10,000 sites and plot the integrated local density of states (LDOS) $D(\varepsilon)$ for 10 sites closest to the centre of the lattice, as shown in Fig.~\ref{fig:lattice}. We need to make sure to include the same number of A and B sublattice sites when calculating the LDOS, since for higher LL the wavefunctions have different amplitudes on both sublattices. The LDOS is
\begin{equation}
	D(\varepsilon)=\sum_n \Theta(\varepsilon-\varepsilon_n)\sum_{i=1}^{10} |\psi_n(\vec x_i)|^2
\end{equation}
where the first sum is over all eigenvalues $\varepsilon_n$ with corresponding normalized eigenfunction $\psi_n(\vec x)$ and the second sum is over the 10 central sites at positions $\vec x_i$. $\Theta(x)$ denotes the Heaviside step function.
We use the strain profile \eqref{eq:strain_profile}.

\begin{figure}
	\centering
	\includegraphics[width=0.5\columnwidth]{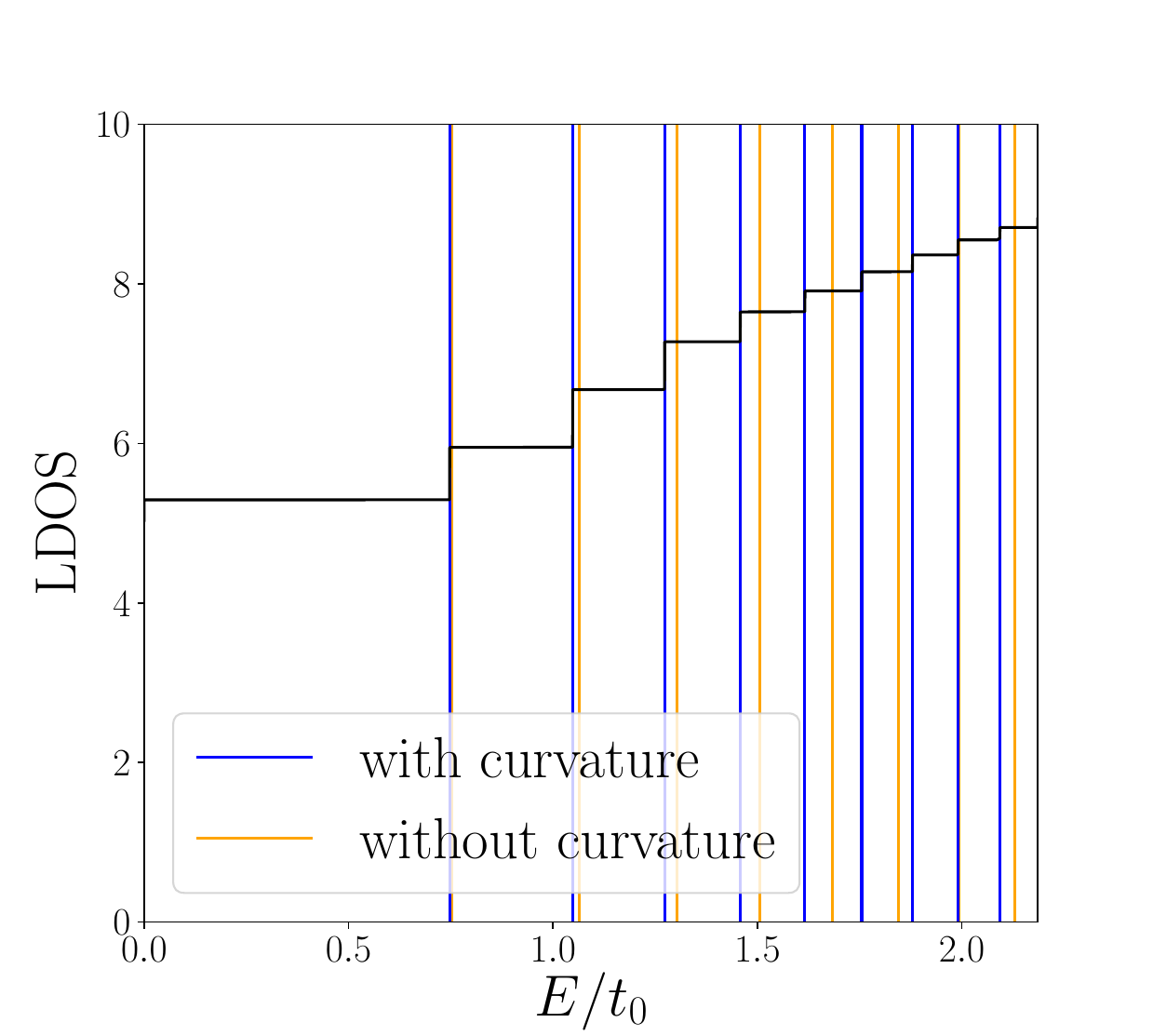}
	\caption{Tight-binding results for the strain profile \eqref{eq:strain_profile} with TB parameter $\beta=3$. We perform exact diagonalization (ED) of a lattice with 10,000 sites. We show the integrated LDOS for the parameter $a u_B/L=0.0105$. We plot the expected position of the pseudo Landau levels with and without the additional contribution from the curvature. It can clearly be seen that the curvature term is important in order to match the field theory result to the ED result. }
	\label{fig:10LL}
\end{figure}

\begin{figure}
	\centering
	\includegraphics[width=0.5\columnwidth]{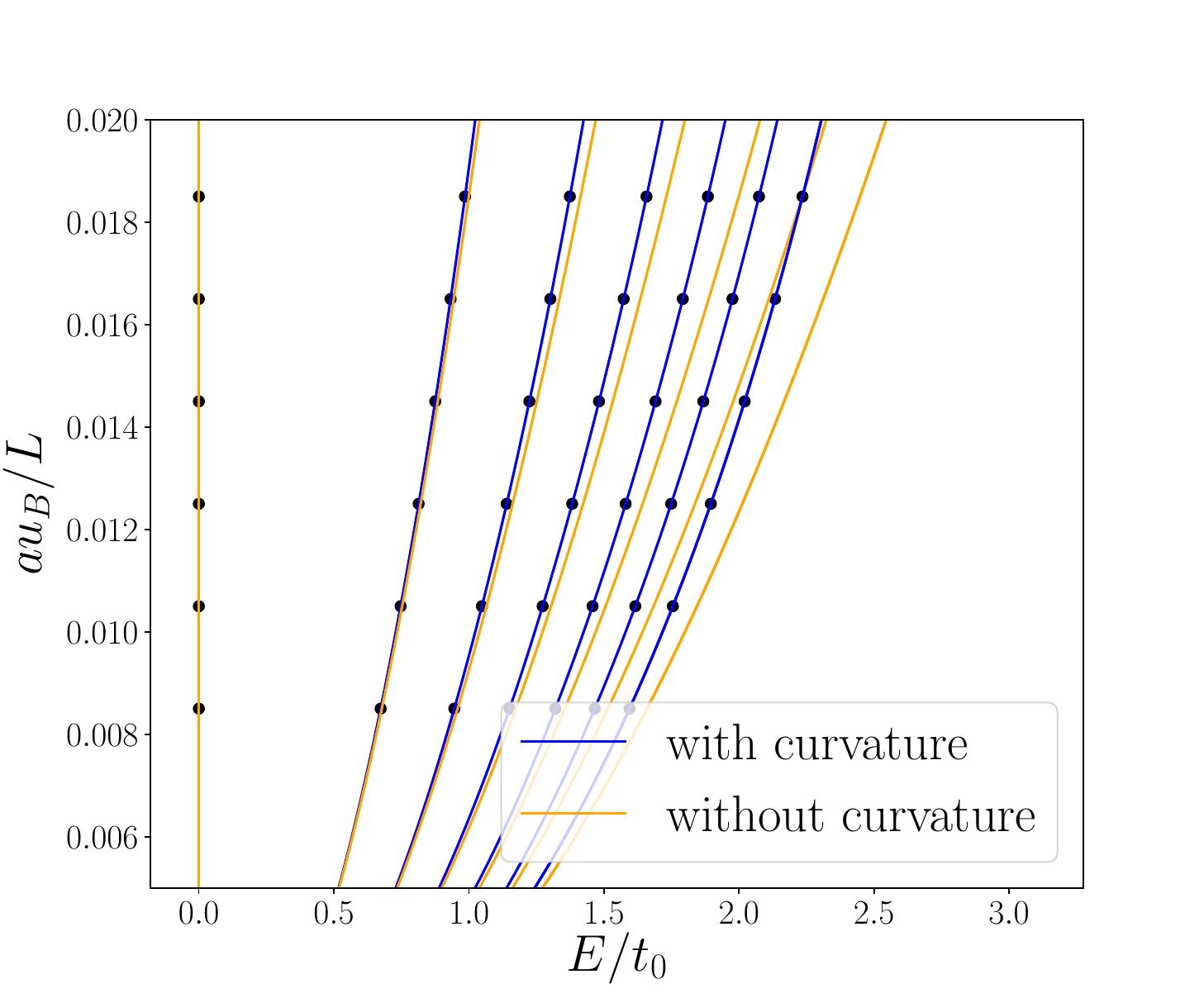}
	\caption{Tight-binding results for the strain profile \eqref{eq:strain_profile} with TB parameter $\beta=3$. We perform exact diagonalization (ED) of a lattice with 10,000 sites. The circles show the location of the pseudo Landau levels as determined by the peaks in the local density of states (LDOS) as a function of the applied strain $a u_B/L$. We compare the prediction \eqref{eq:energy_levels} with (blue curve) and without curvature (orange curve) to the data and see that only the model with curvature accurately fits the data.}
	\label{fig:comparison}
\end{figure}

In Fig.~\ref{fig:10LL} we plot the LDOS and compare to the expression \eqref{eq:energy_levels} with $B$ and $\cK$ given by \eqref{eq:result_for_B} and \eqref{eq:result_for_K} respectively. The spectrum shows clear jumps at the energies corresponding to the pseudo Landau levels. We see that we need to include the effect of the non-zero curvature in order to accurately fit the positions of the pseudo Landau levels. We note that the zeroth pseudo LL is the sharpest, the reason being that the existence of a certain number of zero-energy states is guaranteed by an index theorem, which does not rely on the magnetic field being uniform \cite{index}. Higher pseudo LLs can be smeared out since the small momentum expansion that we used to derive the field theory is no longer valid \cite{Goerbig_Review}. In Fig.~\ref{fig:comparison} we compare the positions of the pseudo LLs with the predictions with and without curvature for different values of $u_B$ which shows that the prediction with curvature matches the numerical data for all $u_B$ investigated.

An additional feature of the QHE in constant curvature is that the pseudo LL degeneracy now depends on the LL index $n$ due to the Wen-Zee shift \cite{WenZee,Pnueli:1994}. Unfortunately the LDOS is not a good probe for this effect and we do not see this effect in our numerics. The Wen-Zee shift is a global property. One can derive the LL degeneracy using the Index Theorem, which requires an integral over the entire manifold \cite{Pnueli:1994}. In addition, the derivation of the Wen-Zee shift uses the assumption that we have a sphere with a constant curvature, which we do not have here. We only have a manifold with approximately constant curvature near the center. So in principle, one should not expect to obtain Wen-Zee shift from LDOS of the current setup.

\section{Comparison to previous work}\label{sec:comparison}

In this section we consider how our results compare with other works that have addressed the same problem, particularly in light of the different subtleties in the calculation that we have discussed throughout.

First, regarding the order of expansion in strain $u_{ij}$, most works like Refs. \cite{Juan:2012,Juan:2013,First_order} initially considered only an expansion to linear order. Later works like \cite{Mazir2013,Oliva_Leyva_2017} extended the expansion of the Hamiltonian to second order in strain, obtaining expressions for the gauge field and Fermi velocity \cite{Oliva_Leyva_2017}. These works however focused only on the computation of the effective magnetic field, without drawing the connection to curved spaces and without computing the Riemann curvature. This is of key importance since the Riemann curvature only has a contribution to second order in strain for the profile we considered. If any of the previous works had computed the lattice spectrum in the presence of strain that leads to spatially constant B, the LL energies would not have matched those predicted by the continuum calculations, because none of them considered the effect of the curvature. 

A related issue is that of the order of derivatives of the strain field that needs to be included. Ref. \cite{Juan:2012} and several subsequent works did not consider derivatives of the strain field in the expansion, but in this paper we do (Eqs. \eqref{eq:vb} and \eqref{eq:At}). The difference in the magnetic field and curvature that results from including this does not appear to first order in strain, but these corrections become important for the terms to second order in strain. Again, calculations ignoring these effect would not have resulted in a perfect match with numerics. This is further explained in Appendix \ref{sec:App_2nd_order_lin}.

Another key point in the matching with continuum calculations is to ensure that the field normalization is the same as that inherited from a TB model, which requires the rescaling of the field \eqref{eq:rename}. The inclusion of this is a difference with Refs. \cite{Khaidukov2016,Zubkov2}, for example. This rescaling has been explicitly used, for the mentioned reason in Ref. \cite{Kozlovsky20} in the context of topological insulator surface states.    

Another source of difference with other works is the question of whether to perform the momentum expansion around the original Dirac points of the honeycomb lattice, or around new, shifted Dirac points that are a function of the strain, as done in \cite{Zubkov2}. In Appendix \ref{sec:A_K_pt} we show that the second choice is ill-defined when carried out to second and higher orders in strain, while the first choice leads to the effective theory that matches the numerical calculation precisely.

Finally, the same strain profile that we used is also studied in \cite{Khaidukov2016}. Their expression for the energy levels differs from our result in two regards. On the one hand they employ a different definition of the emergent magnetic field, namely $\vec H=\nabla\times\vec A$, whereas we use the curved space formula $\vec B=(\nabla\times\vec A)/\sqrt{g}$. This accounts for the apparent difference between our result and that of \cite{Khaidukov2016}, when written directly in terms of the strain field the results are the same, except that \cite{Khaidukov2016} neglects again the curvature correction to the energy levels (the $n^2$ term).  

\section{Conclusions}

In this work, we have shown that the effective low-energy theory governing strained graphene is that of a Dirac fermion in curved space coupled to an artificial gauge field. Armed with the precise mapping between lattice and continuum field theory formulations, we aimed to test our theory with the simple prediction of the Quantum Hall effect in constant curvature. We have been able to reproduce the well known spectrum of this problem in exact diagonalization of the real space tight-binding Hamiltonian and the agreement between the theory and the numerics vindicates out low-energy theory. This is the first term that these corrections to the Landau levels due to curvature have been seen in strained graphene. 

Throughout the paper, we have commented on several subtle points in the derivation of the mentioned mapping that are important to obtain our results. First, the volume form must be absorbed in the continuum fields in order to define a Hamiltonian for Dirac fermions in curved space that can be compared with a lattice calculation. Second, we have addressed a controversy regarding the origin of the momentum expansion in the tight-binding Hamiltonian. And third, we have carried out the calculation to second order in strain when matching with the constant curvature scenario. The importance of these three points is illustrated simply by realizing that if these are not taken into account, the match with numerics is not obtained. 

Besides straining ordinary graphene, there are two other promising platforms that offer more control over the desired strain profile. The first is a photonic lattice, where waveguides are etched into a crystal in a hexagonal arrangement and hopping of photons between waveguides is well-described by a tight-binding model. Signatures of the quantum Hall effect in the photonic analogue of strained graphene have already been observed \cite{Rechtsman1,Rechtsman2,Rechtsman3,Rechtsman4}. The main signature seen so far in the photonic analogue are the robust chiral edge modes. It would interesting to extend this work in order to be able to measure the Landau levels themselves. However, this would require measuring the local density of states, something that has proved elusive in this system so far. On the other hand, a system which offers a way of measuring energies directly is the sonic lattice \cite{SonicLattice,Sonic2,Sonic3}. A similar Hamiltonian has also been constructed for ultracold atoms in an optical lattice \cite{Ultracold_atoms}. Finally, another avenue for further research would be to consider different strain profiles that may be able to simulate more exotic gravitational analogues, for example the graphene analogue of a wormhole \cite{Wormholes}. 

\section*{Acknowledgements}
	
F.J. especially thanks Joel E. Moore for insightful discussions on this topic. D.X.N would like to thank Paul Wiegmann and Dam Thanh Son for for a discussion on the UV completeness of the effective theory. The authors also thank Carlos Hoyos, Steve Simon and Christopher Herzog for comments on a previous version of this manuscript. The calculation of the curvature from the metric was performed using the Mathematica package `Riemannian Geometry and Tensor Calculus' (RGTC) \cite{RGTC}. This work was supported by grants EP/N01930X/1 and EP/S020527/1. DXN was supported partially by Brown Theoretical Physics Center.  

\appendix
\section{$2+1 D$ Dirac fermion in static curved space }
\label{sec:DetailDirac}
\subsection{Action in curved space-time }
We recall the action \eqref{eq:LagNonHerm} of a spin-$\frac{1}{2}$ Dirac fermion in curved space-time
\begin{equation}
	\label{eq:Lagcur0}
	\cS=i\int d^3 x \sqrt{|g|}\bar{\Psi} \gamma^\mu(\partial_\mu-iA_\mu-\frac{i}{2}\omega_\mu^{\alpha \beta}\sigma_{\alpha\beta})\Psi.
\end{equation}
The Hermitian conjugate of \eqref{eq:Lagcur0} is
\begin{equation}
	\cS^\dagger=i\int d^3x \sqrt{|g|} e^\mu_\alpha [-\d_\mu \Psi^\dagger (\upgamma^\alpha)^\dagger-i\Psi^\dagger A_\mu (\upgamma^\alpha)^\dagger-\frac{i}{2}\Psi^\dagger\omega^{\beta \gamma}_\mu (\sigma_{\beta\gamma})^\dagger(\upgamma^\alpha)^\dagger ](\upgamma^0)^\dagger \Psi.
\end{equation}
We then use the following identities 
\begin{equation}
	(\upgamma^0)^\dagger=\upgamma^0, \quad (\upgamma^\alpha)^\dagger\upgamma^0=\upgamma^0\upgamma^\alpha
\end{equation}
together with the definition of $\sigma_{\alpha\beta}$ and rewrite the action $\cS^\dagger$ as 
\begin{equation}
	\label{eq:Lagcur0dg}
	\cS^\dagger=i\int d^3 x \sqrt{|g|}\bar{\Psi} \left[\gamma^\mu(-\overleftarrow{\partial_\mu}-iA_\mu)-\frac{i}{2}\omega_\mu^{\alpha \beta}\sigma_{\alpha\beta}\gamma^\mu \right]\Psi,
\end{equation}
where $\overleftarrow{\partial_\mu}$ only acts on $\bar{\Psi}$ respectively. We now will show that the actions \eqref{eq:Lagcur0} and \eqref{eq:Lagcur0Herm} are equivalent up to a surface term. Define $\cS'=\frac{1}{2}\left(\cS+\cS^\dagger\right)$. Then from \eqref{eq:Lagcur0} and \eqref{eq:Lagcur0dg}, we have 
\begin{align}
	\label{eq:diff}
	\cS-\cS'&=\frac{i}{2}\int d^3 x \sqrt{|g|}\bar{\Psi}\left(\gamma^\mu\overrightarrow{\nabla}_\mu+\overleftarrow{\nabla}_\mu \gamma^\mu \right)\Psi,
\end{align}
where the covariant derivatives $\overrightarrow{\nabla}_\mu, \overleftarrow{\nabla}_\mu$, which are defined as 
\begin{equation}
	\overrightarrow{\nabla}_\mu =\overrightarrow{\partial}_\mu-\frac{i}{2}\omega_\mu^{\alpha\beta}\sigma_{\alpha\beta}, \qquad \overleftarrow{\nabla}_\mu =\overleftarrow{\partial}_\mu+\frac{i}{2}\omega_\mu^{\alpha\beta}\sigma_{\alpha\beta}, 
\end{equation}
only act on $\Psi$ and $\bar{\Psi}$. 
We use the expression of the covariant derivative of the gamma matrices \cite{Freedman:SUGRA}
\begin{equation}
	\nabla_\mu \gamma_\nu=\d_\mu \gamma_\nu-\frac{i}{4}\omega_\mu^{\alpha\beta}\left[\sigma_{\alpha\beta},\gamma_\nu\right]-\Gamma^{\rho}_{\mu\nu}\gamma_\rho,
\end{equation}
which implies that $\gamma_\mu$ is not only a covariant vector with index $\mu$ but also has two spinor indices which need to be taken care of properly in the definition of the covariant derivative. Using the anti-commutation relation of gamma matrices and the definition \eqref{eq:defgamma}, we have 
\begin{equation}
	\nabla_\mu \gamma_\nu=\upgamma^\alpha\left(\d_\mu e_{\alpha\nu}+\omega_{\mu\alpha\beta}e^\beta_\nu-\Gamma^{\rho}_{\mu\nu}e_{\alpha\rho}\right).
\end{equation}
The right hand side of the above equation vanishes due to the tetrad postulate \eqref{eq:tetrad}, we then can rewrite equation \eqref{eq:diff} as 
\begin{align}
	\cS-\cS'&=\frac{i}{2}\int d^3x \sqrt{|g|}\nabla_\mu \left(\bar{\Psi}\gamma^\mu \Psi\right)\\&=\frac{i}{2}\int d^3x \sqrt{|g|}\left[\d_\mu \left(\bar{\Psi}\gamma^\mu \Psi\right)+\Gamma^{\mu}_{\mu\nu}\left(\bar{\Psi}\gamma^\nu \Psi\right)\right],
\end{align}
where we recognize that the right hand side is the covariant derivative of the current density operator. We then use the identity
$
\Gamma^\mu_{\mu\nu}=\frac{\d \ln(|g|)}{\d x^\nu}
$
to show that the above equation is just a surface term
\begin{equation}
	\label{eq:total_derivative}
	\cS-\cS'=\frac{i}{2}\int d^3x \d_\mu \left(\sqrt{|g|}\bar{\Psi}\gamma^\mu \Psi\right).
\end{equation}
So indeed we confirmed that $\cS$ and $\cS'$ are equivalent up to a surface term and we can use either of them for the theory of a $2+1D$ Dirac fermion in curved space-time. 

Furthermore, in a system with boundaries in the spatial directions, the natural boundary conditions should be chosen such as Dirac particle can't escape the system. Such boundary conditions satisfy the following constraint
\begin{equation}
\label{eq:bc1}
	\hat{n}_i \left(\sqrt{|g|}\bar{\Psi}\gamma^i \Psi\right)= \hat{n}_i J^i=0
\end{equation}
where $\hat{n}$ is the normal vector of the boundary and
\begin{equation}
	J^i=\sqrt{|g|}\bar{\Psi}\gamma^i \Psi
\end{equation}
is the current operator in curved space. In addition, if one considers the boundary in the time direction, the vanishing of \eqref{eq:total_derivative} is the requirement for conservation of the global charge $d \mathcal{Q}/dt=0$ with 
\begin{equation}
	\mathcal{Q}=\int d^2 \mathbf{x}\, J^0, \qquad J^0=\sqrt{|g|}\bar{\Psi}\gamma^0\Psi=\sqrt{|g|}\Psi^\dagger\Psi.
\end{equation}
 Thus the boundary term \eqref{eq:total_derivative} vanishes with a suitable choice of the boundary conditions. In other words, the actions $\mathcal{S}$ and $\mathcal{S}'$ are equivalent in any physical system.
\subsection{Hamiltonian in static curved space}
Combining equations \eqref{eq:Lagcur0} and \eqref{eq:Lagcur0dg}, we obtain 
\begin{equation}
	\label{eq:Lagcur1}
	\cS'=i\int d^3 x \sqrt{|g|}\bar{\Psi} \left[\gamma^\mu(\overleftrightarrow{\partial}_\mu-iA_\mu)+\frac{i}{4}\omega_\mu^{\alpha \beta}\{\sigma_{\alpha\beta},\gamma^\rho\}e_{\rho}^\mu\right]\Psi,
\end{equation}
where $\overleftrightarrow{\partial}_\mu=\frac{1}{2}\left(\overrightarrow{\d}_\mu-\overleftarrow{\d}_\mu\right)$ only acts on fermion fields. From the definition of the static curved space vielbein, we see that the anti-commutator in the above equation vanishes. Then in static curved space, we obtain the action 
\begin{equation}
	\label{eq:Lagcur2}
	\cS'=i\int d^3 x \sqrt{\hat{g}}\bar{\Psi} \left[\gamma^\mu(\overleftrightarrow{\partial}_\mu-iA_\mu)\right]\Psi,
\end{equation}
with corresponding Hermitian Hamiltonian
\begin{equation}
	H=-i\int d^2 x \sqrt{\hat{g}}\left[\Psi^\dagger e^i_a \sigma^a\left(\overleftrightarrow{\partial}_i-iA_i\right)\Psi-iA_0\Psi^\dagger\Psi\right],
\end{equation}
where $\hat{g}=\det(g_{ij})$. If we consider the Dirac action with the Fermi velocity $v_F$ replacing the speed of light,
\begin{equation}
	\cS=i\int d^3 x \sqrt{|g|}\bar{\Psi} \left[\gamma^0(\overrightarrow{\nabla}_0-iA_0)+v_F\gamma^i(\overrightarrow{\nabla}_i-iA_i)\right]\Psi,
\end{equation}
one can repeat the calculation in the previous section with slight modifications and obtain the Hermitian Hamiltonian in static curved space
\begin{equation}
\label{eq:HamGen}
	H=-i\int d^2 x \sqrt{\hat{g}}\left[\Psi^\dagger v_F e^i_a \sigma^a\left(\overleftrightarrow{\partial}_i-iA_i\right)\Psi-iA_0\Psi^\dagger\Psi\right],
\end{equation}
If we consider an applied magnetic field only and chose the Coulomb gauge $A_0=0$, the Hamiltonian has the form
\begin{equation}
	H=-iv_F\int d^2 x \sqrt{\hat{g}}\Psi^\dagger e^i_a \sigma^a\left(\overleftrightarrow{\partial}_i-iA_i\right)\Psi.
\end{equation}
The above equation is nothing but the Hamiltonian \eqref{eq:Hamcur0} in the main text.  

Now we showed in \eqref{eq:total_derivative} that $\mathcal S$ and $\mathcal S'$ differ by a total derivative, hence they should give rise to the same equation of motion for the field $\Psi$. How does this reconcile with the fact that the action $\mathcal S$ in \eqref{eq:Lagcur0} has a spin connection term while the action $\mathcal{S}'$ (valid for static curved space) in \eqref{eq:Lagcur2} does not? When we compute the equation of motion for $\Psi$ from \eqref{eq:Lagcur2}, we need to integrate by parts to convert the derivative $\overleftrightarrow{\partial}_\mu$ into a derivative $\d_\mu$ acting only on the right. In this process, we end up with a derivative of $\sqrt{\hat{g}}$ and a derivative of the vielbein (since the vielbein is implicit in $\gamma^\mu$). The term including the derivative of $\sqrt{\hat{g}}$ is cancelled after we redefine the Dirac field as in equation \eqref{eq:rename}. The derivative of the spin connection term can be shown to be precisely equivalent to the term coming from the spin connection when the equation of motion is derived from \eqref{eq:Lagcur0}
\footnote{We note that in \cite{Juan:2012}, there is a term with the derivative the space-dependent Fermi velocity in the equation of motion, which corresponds to the derivative of the vielbein after the mapping we will introduce in section \ref{sec:map}.}.

\section{Derivation of effective tight-binding Hamiltonian}
\label{sec:TBApx}
The TB Hamiltonian \eqref{eq:TB0} is Hermitian by construction. We will map the final result to the Hamiltonian of a $2+1D$ Dirac fermion in static curved space. Since the Hamiltonian is Hermitian, the corresponding action is $\cS'$ instead of $\cS$, and one then discover that the mapped Hamiltonian does not have the spin-connection term which is \eqref{eq:Hamcur}. In the continuum limit, we define the field operator of sub-lattice $A$ and $B$ as well as their Fourier modes:
\begin{equation}
	\label{eq:wfF}
	\psi_I(\mathbf{x})=\sum_{\mathbf{k}} \frac{1}{N}  e^{-i \mathbf{k}\x}\psi_I(\mathbf{k}), \quad \psi_I(\mathbf{k})=\sum_{\x} e^{i \mathbf{k}\x}\psi_I(\mathbf{x}). 
\end{equation}
where $I=A,B$ with the normalization
\begin{equation}
	\sum_{\x}e^{i\mathbf{k}\x}=N\delta_{\mk,0},\qquad \sum_{\mk}e^{i\mathbf{k}\x}=N\delta_{\x,0}, 
\end{equation}
We can rewrite the Hamiltonian \eqref{eq:TB0} as 
\begin{equation}
	H^{TB}=\sum_{n,\mR_i}\sum_{\mk,\mk'}\frac{1}{N^2}t_n(\mR_i)\bigg(\psi^{\dagger}_A(\mk)\psi_B(\mk')e^{i\mk \mR_i - i \mk'(\mR_i+\ml_n)}+\psi^\dagger_B(\mk')\psi_A(\mk)e^{-i\mk \mR_i + i \mk'(\mR_i+\ml_n)}\bigg)
\end{equation}
we redefine $\mk \leftrightarrow \mk'$ on the second term and define 
\begin{equation}
	\psi(\mk)=\begin{pmatrix}
		\psi_A(\mk)\\ \psi_B(\mk) 
	\end{pmatrix}
\end{equation}
to obtain the matrix form equation 
\begin{equation}
	\label{eq:HTB}
	H^{TB}=\sum_{n,\mk,\mk'}\frac{1}{N^2}t_n(\mk-\mk')\psi^\dagger(\mk)\begin{pmatrix}
		0 &e^{-i\mk' \ml_n} \\
		e^{i \mk \ml_n}& 0
	\end{pmatrix}\psi(\mk'),
\end{equation}
with the definition of Fourier transformation 
\begin{equation}
	t_n(\mathbf{x})=\sum_{\mathbf{k}} \frac{1}{N}  e^{-i \mathbf{k}\x}t_n(\mathbf{k}), \qquad t_n(\mathbf{k})=\sum_{\x} e^{i \mathbf{k}\x}t_n(\mathbf{x}). 
\end{equation}
\subsection{Expansion around the $\mK$ point}
We define the two Dirac points $\mK=(\frac{4\pi}{3\sqrt{3}a},0)$ and $\mK'=-\mK$.
We then define $\mk=\mq+\mK$, and redefine $\psi(\mK+\mq)\rightarrow \psi(\mq)$. Expanding up to second order in the small momenta $\vec q$ and $\vec q'$, we find
\begin{equation}
	H^{TB}=i\sum_{n,\mq,\mq'}\frac{1}{N^2}t_n(\mq-\mq')\psi^\dagger(\mq)\frac{\vec{\sigma}\cdot\vec{l}_n}{a}\sigma^3 \times\begin{pmatrix}
		1+i q_i l^i_n-\frac{1}{2}q_iq_jl_n^il_n^j & 0 \\
		0 & 1-i q'_i l^i_n-\frac{1}{2}q_i'q_j'l_n^il_n^j
	\end{pmatrix}\psi(\mq')
\end{equation}
we then define 
\begin{equation}
	\label{eq:k_transf}
	\mathbf{Q}=\frac{1}{2}(\mq+\mq'), \qquad \ms=\frac{1}{2} (\mq-\mq').
\end{equation}
We can rewrite the above Hamiltonian in the following form

	\begin{multline}
		H^{TB}=i\sum_{n,\mQ,\ms}\frac{1}{4 N^2}t_n(2\ms)
		\psi^\dagger(\mQ+\ms)\frac{\vec{\sigma}\cdot\vec{l}_n}{a}\sigma^3 \\\times \begin{pmatrix}
			1+i(Q_i+s_i) l^i_n-\frac{1}{2} (Q_i+s_i)(Q_j+s_j) l_n^il_n^j& 0 \\
			0 & 1-i(Q_i-s_i) l^i_n-\frac{1}{2} (Q_i-s_i)(Q_j-s_j) l_n^il_n^j
		\end{pmatrix}\psi(\mQ-\ms),
	\end{multline}
	
	where the factor $1/4$ comes from the change of variable (Jacobian). Using the Fourier transformation to convert it back to the coordinate space, we obtain
	\small{
	\begin{multline}
		\label{eq:HamTB1}
		H^{TB}=i\sum_{n,\mQ,\ms,\x,\y}\frac{1}{4 N^2}t_n(2\ms)e^{-i\mQ (\x-\y)-i\ms(\x+\y)}\times\\
		\psi^\dagger(\x)\frac{\vec{\sigma}\cdot\vec{l}_n}{a}\sigma^3 \begin{pmatrix}
			1+i(Q_i+s_i) l^i_n-\frac{1}{2} (Q_i+s_i)(Q_j+s_j) l_n^il_n^j& 0 \\
			0 & 1-i(Q_i-s_i) l^i_n-\frac{1}{2} (Q_i-s_i)(Q_j-s_j) l_n^il_n^j
		\end{pmatrix}\psi(\y).
	\end{multline}}
	Using the identity
	$
	\sum_{\mQ}e^{-i\mQ(\x-\y)}=4 N \delta_{\mx-\my,0},
	$
	and the identities
	\begin{equation}
		\sum_{\x,\y}i Q_i e^{-i\mQ (\x-\y)-i\ms(\x+\y)}\psi^\dagger(\x)\Sigma\psi(\my)=\sum_{\x,\y}e^{-i\mQ (\x-\y)-i\ms(\x+\y)}\frac{1}{2}\left(\frac{\d \psi^\dagger(\mx)}{\d x^i}\Sigma\psi(\y)-\psi^\dagger(\x)\Sigma\frac{\d \psi(\y) }{\d y^i} \right),
	\end{equation}
	\begin{equation}
		\sum_{\x,\my}i s_i e^{-i\mQ (\x-\y)-i\ms(\x+\y)}\psi^\dagger(\x)\Sigma\psi(\my)=\sum_{\x,\y}e^{-i\mQ (\x-\y)-i\ms(\x+\y)}\frac{1}{2}\left(\frac{\d \psi^\dagger(\mx)}{\d x^i}\Sigma\psi(\y)+\psi^\dagger(\x)\Sigma\frac{\d \psi(\y) }{\d y^i} \right)
	\end{equation}

up to surface terms, where $\Sigma$ is any $2\times 2$ matrix, we go back to position space, the results \eqref{eq:HamCurK} together with \eqref{eq:vb} and \eqref{eq:At} follow after we replace 
\begin{equation}
	\sum_{\mx}\rightarrow \frac{1}{a^2} \int d^2 \mx
\end{equation}
and integrate by parts. We discard all terms that have more than one derivative acting on fermion fields, since these cannot be brought into the form of the Dirac Hamiltonian. We also discard terms that have more than one derivative acting on the hopping $t_n(\x)$ and terms with higher order in $\overleftrightarrow{\partial}$. There terms are all higher-order in the momentum and will be relevant far from the Dirac points, ie once we get to high Landau levels.

\subsection{Expansion around the $\mK'$ point}
Repeating the same calculations in the above subsection, we obtain the effective Hamiltonian
\begin{align}
	\label{eq:HamCurK'}
	H^{TB}=&\int d^2 \x v_F\psi'^\dagger(\x)\left(i\sigma_i \tilde{v}^{ij}(\x)\overleftrightarrow{\d}_j -\sigma^i A^{s}_i (\x) \right)\psi'(\x)
\end{align} 
where $\tilde{v}^{ij}(\x)$ and $A_i(\x)$ are the same as in \eqref{eq:vb} and \eqref{eq:At}. The definition of $\psi'(\x)$ is 
\begin{equation}
	\label{eq:psiK'}
	\psi'(\mx)=e^{-i \mK \x}\begin{pmatrix}
		\psi_B(\mx)\\ \psi_A(\mx) 
	\end{pmatrix}
\end{equation}
Notice again the minus sign in the derivative term of \eqref{eq:HamCurK'} in comparison with \eqref{eq:HamCurK}. 
\subsection{Valley dual transformation}
From \eqref{eq:HamCurK} and \eqref{eq:HamCurK'} we derive the Schr\"odinger equation of the effective theory near $\mK$ \begin{equation}
	\label{eq:SchK}
	-v_F\sigma_i\left(i\tilde{v}^{ij}(\x)\d_j +\frac{i}{2}\partial_j\tilde{v}^{ij}(\x)+ A^{s}_i (\x) \right)\psi(\x)=E \psi(\x),	
\end{equation}
and near the $\mK'$ point
\begin{equation}
	\label{eq:SchK'}
	v_F\sigma_i \left(i\tilde{v}^{ij}(\x)\d_j +\frac{i}{2}\partial_j\tilde{v}^{ij}(\x)- A^{s}_i (\x) \right)\psi'(\x)=E \psi'(\x)	.
\end{equation}
We take the complex conjugate of equation \eqref{eq:SchK'} then multiply by $\sigma^1$ and use the identity
\begin{equation}
	\sigma^1(\sigma_i)^*=\sigma_i\sigma^1 \qquad (i=1,2),
\end{equation}
to obtain
\begin{equation}
	\label{eq:SchK'ph}
	-v_F\sigma_i\left(i \tilde{v}^{ij}(\x)\d_j +\frac{i}{2}\partial_j\tilde{v}^{ij}(\x)+A^{s}_i (\x) \right)\sigma^1\psi'^*(\x)=E \sigma^1\psi'^*(\x),	
\end{equation}
which is the same as the Schr\"odinger equation of the effective theory near the $\mK$ point  \eqref{eq:SchK}. From the above transformation, we see that each eigenenergy of the effective Hamiltonian near $\mK'$ corresponds to the same eigenenergy of the effective Hamiltonian near $\mK$. We also see that the transformation of the wave-function has the form of a valley dual (VD) transformation in $2+1D$ \footnote{It looks similar to the particle-hole (PH) transformation in \cite{Son:Dirac}. However, in our case, we only have the transformation of the wave-function instead of the field operator as in \cite{Son:Dirac}. As a consequence, a particle field near $\mK$ maps to a particle field near $\mK'$.}  
\begin{equation}
	\psi'(x) \xrightarrow{VD} \sigma^1 \psi'^*(\mx).
\end{equation}
Using equation \eqref{eq:psiK'}, we see that under the P-H transformation, we obtain 

\begin{equation}
	\psi'(x) \xrightarrow{VD} e^{i \mK \x}\begin{pmatrix}
		\psi^*_A(\mx)\\ \psi^*_B(\mx) \end{pmatrix}.
\end{equation}
The VD transformations transform the effective theory near $\mK$ into the effective field theory near $\mK'$. 

\section{Which $\vec K$ point should one expand about?}
\label{sec:A_K_pt}
In this appendix, we analyze the expanding around space-dependent $\mathbf{K}$ point apprach in Ref \cite{Zubkov:2015}. We will show that by careful calculation, this approach is ill-defined at the higher order in strain field. The starting point is---as before---\eqref{eq:HTB}
\begin{equation}
	H^{TB}=\sum_{n,\mk,\mk'}\frac{1}{N^2}t_n(\mk-\mk')\psi^\dagger(\mk)\begin{pmatrix}
		0 &e^{-i\mk' \ml_n} \\
		e^{i \mk \ml_n}& 0
	\end{pmatrix}\psi(\mk'),
\end{equation}
and we define the new variables
\begin{equation}
	\mathbf{Q}=\frac{1}{2}(\mk+\mk'), \qquad \ms=\frac{1}{2} (\mk-\mk').
\end{equation}
Beware that this definition is not the same as \eqref{eq:k_transf} due to the difference between $\vec k$ and $\vec q$. Remembering the Jacobian for this transformation, we obtain 
\begin{equation}
	H^{TB}=\sum_{n,\vec s,\vec Q}\frac{1}{4N^2}t_n(2\vec s)\psi^\dagger(\mQ+\ms)\begin{pmatrix}
		0 &e^{-i(\mQ-\ms) \ml_n} \\
		e^{i (\mQ+\ms) \ml_n}& 0
	\end{pmatrix}\psi(\mQ-\ms),
\end{equation}
Now define 
\begin{equation}
	\label{eq:Hexpr}
	H^{TB}=\sum_{\vec s,\vec Q}\frac{1}{4N^2}\psi^\dagger(\mQ+\ms)
	\hat H^{TB}(\vec Q,\vec s)\psi(\mQ-\ms),
\end{equation}
where
\begin{equation}
	\hat H^{TB}(\vec Q,\vec s)=\sum_{n}t_n(2\vec s)\begin{pmatrix}
		0 &e^{-i(\mQ-\ms) \ml_n} \\
		e^{i (\mQ+\ms) \ml_n}& 0
	\end{pmatrix}
\end{equation}
We now follow Ref.~\cite{Zubkov:2015} and define the spatially-dependent $\vec K$-point via
\begin{equation}
	\sum_{n}t_n(2\vec s)\begin{pmatrix}
		0 &e^{-i\mK^\pm(2\vec s) \ml_n} \\
		e^{i \mK^\pm(2\vec s) \ml_n}& 0
	\end{pmatrix}=0
\end{equation}
We then expand $\hat H^{TB}(\vec Q,\vec s)$ around $(\vec K^\pm,\vec 0)$. 
\begin{equation}\hat H^{TB}(\vec Q,\vec s)=i \sigma^{3}[(\mp \sigma^{2} \mathbf{f}_{2}(2\vec s)+\sigma^{1} \mathbf{f}_{1}(2\vec s))(\mathbf{Q}-K^{\pm}(2\vec s))-(\sigma^{1} \mathbf{f}_{1}(2\vec s) \mp \sigma^{2} \mathbf{f}_{2}(2\vec s)) \sigma^{3} \mathbf{s}]\end{equation}
where the expressions for $\mathbf{f}_{1,2}(2\vec s)$ and $K^{\pm}(2\vec s)$ can be found in Ref. \cite{Zubkov:2015}. Let us write $\vec K^\pm(2\vec s)=\vec K^{\pm (0)}+\delta\vec K^\pm(2\vec s)$ and consider the term at  second order in strain 
\begin{equation}
	-i\sigma^3\sigma^1\vec f_1(2\vec s)\delta\vec K^\pm(2\vec s)\in \hat H^{TB}(\vec Q,\vec s)
\end{equation}
Now plugging back into \eqref{eq:Hexpr}, we need to evaluate a term of the form 
\begin{equation}
	\sum_{\vec s,\vec Q}\psi^\dagger(\mQ+\ms)
	\vec f_1(2\vec s)\delta\vec K^\pm(2\vec s)\psi(\mQ-\ms)
\end{equation}
Now we transform back to real space 
\begin{equation}
	\sum_{\mx,\my}\sum_{\vec s,\vec Q}\psi^\dagger(\vec x)e^{-i(\mQ+\ms)\vec x}
	\vec f_1(2\vec s)\delta\vec K^\pm(2\vec s)\psi(\my)e^{i(\mQ-\ms)\vec y}
\end{equation}
an evaluating the sum over $\vec Q$ this becomes
\begin{equation}
	N\sum_{\mx}\sum_{\vec s}\psi^\dagger(\vec x)e^{-2i\ms\vec x}
	\vec f_1(2\vec s)\delta\vec K^\pm(2\vec s)\psi(\mx)
\end{equation}
In particular let us study the matrix 
\begin{equation}
	\label{eq:sing}
	\sum_{\vec s}e^{-2i\ms\vec x}
	\vec f_1(2\vec s)\delta\vec K^\pm(2\vec s)=\sum_{\ms,\my,\my'}e^{-2i\ms\vec x}e^{2i\my\ms+2i\my'\ms}
	\vec f_1(\vec y)\delta\vec K^\pm(\vec y')=\frac{N}{2}\sum_{\my}\vec f_1(\vec y)\delta\vec K^\pm(\vec x-\vec y)
\end{equation}
Now assuming $t_n(\vec x)$ is well-defined, then so will $\vec f_1(\vec x)$ and $\delta\vec K^\pm(\vec x)$ be well-defined.  However, in general, the sum in \eqref{eq:sing} will not converge. This shows that the expansion about the spatially-dependent $\vec K$-point is not well-defined.  One can check that it is the case for our choice of strain profile \eqref{eq:strain_profile}. Defining $u_{ij}\equiv \partial_i u_j$ we have
\begin{equation}
    \vec f_1(\vec x)=v_F\begin{pmatrix}
		1-\beta u_{11}(\mx)\\ -\beta u_{12}(\mx) \end{pmatrix}=v_F\begin{pmatrix}
		1-\frac{2\beta u_B}{L}x_2\\ -\frac{2\beta u_B}{L}x_1 \end{pmatrix}
\end{equation}
and
\begin{equation}
    \delta\vec K^\pm(\vec x)=\pm\frac{\beta}{2a}\begin{pmatrix}
		u_{22}(\mx)-u_{11}(\mx)\\ u_{12}(\mx)+u_{21}(\mx) \end{pmatrix}=\pm\frac{2\beta}{a}\begin{pmatrix}
		-x_2\\ x_1 \end{pmatrix}
\end{equation}
for the given strain profile. Hence the expression \eqref{eq:sing} contains terms like 
\begin{equation}
    N\sum_{\vec y} y_1 (x_1-y_1)
\end{equation}
which do not converge.

\section{Uniform dilatation}
Let us consider the case of a uniform dilatation of our atomic lattice
\begin{equation}
	\vec x'=\vec x+\vec u(\vec x)=\vec x+\varepsilon\vec x
	\label{eq:transf}
\end{equation}
and so with the hopping \eqref{eq:tn_lin} we find
\begin{equation}
	t'=t_0(1-\beta\varepsilon)
\end{equation}
and hence via \eqref{eq:vb} we have 
\begin{equation}
	v_F'=(1-\beta\varepsilon)v_F.
\end{equation}
But now we need to remember that our Dirac Hamiltonian is written in the atomic frame ($\vec x$ coordinates), instead of in the lab frame ($\vec x'$ coordinates). If we start with the Hamiltonian in the atomic frame
\begin{equation}H_{\mathrm{atomic}}=-i v_F'\int d^{2} \vec{x}\ \psi^{\dagger} \sigma^{i} \frac{\partial}{\partial x_{i}} \psi\end{equation}
then in the lab frame we obtain, then by virtue of the transformation \eqref{eq:transf} (see e.~g.~\cite{LeonTBG} for details)
\begin{equation}H_{\mathrm{lab}}=\int d^{2} \vec{x'} \psi^{\dagger}\bigg[-i v_F'\left(\sigma^i+\frac{\partial u_i}{\partial x'_j}\sigma^j\right) \frac{\partial}{\partial x_i'}+v_F'\left(\boldsymbol{K} \cdot \partial'_i \boldsymbol{u}\right)\sigma^i\bigg] \psi\end{equation}
which simplifies to 
\begin{equation}H_{\mathrm{lab}}=v_F\sigma^i\int d^{2} \vec{x'} \psi^{\dagger}\bigg[-i \left(1+\varepsilon-\beta\varepsilon\right) \frac{\partial}{\partial x_i'}+\varepsilon(1-\beta\varepsilon) K_i\bigg] \psi\end{equation}
Now going to momentum space
\begin{equation}H_{\mathrm{lab}}=v_F\sigma^i\frac{1}{N^2}\sum_\vec{k} \psi^{\dagger}(\vec k)\bigg[ \left(1+\varepsilon-\beta\varepsilon\right)(k_i-K_i)+\varepsilon(1-\beta\varepsilon) K_i\bigg] \psi(\vec k)\end{equation}
and up to second order in $\varepsilon$
\begin{equation}H_{\mathrm{lab}}=\frac{v_F\sigma^i}{N^2}\sum_\vec{k} \psi^{\dagger}(\vec k)\bigg[ \left(1+\varepsilon-\beta\varepsilon\right)(k_i-K_i-\delta K_i)\bigg] \psi(\vec k)\end{equation}
where $\delta K_i=-\varepsilon K_i$. So the final result is consistent with our expectations: We have $a\to (1+\varepsilon)a$ and $t\to t(1-\beta\epsilon)$ so $v_F=\frac{3}{2}ta\to v_F(1+\varepsilon-\beta\varepsilon)$ and $K=\frac{4\pi}{3\sqrt{3}a}\to K(1-\varepsilon)$.

\section{Second-order calculation: linear tunneling}
\label{sec:App_2nd_order_lin}
The hopping has the form
\begin{equation}
	t_n(\vec x)=t_0\bigg[1-\frac{\beta}{a^2} \bigg(\vec u(\vec x+\vec l_n)-\vec u(\vec x)\bigg)\cdot \mathbf{l}_n\bigg]
\end{equation}
and
\begin{equation}
	\label{eq:D2}
	\vec{u}(\vec x+\vec l_n)-\vec u(\vec x)\approx (\vec l_n\cdot\nabla) \vec u+\frac{1}{2}(\vec l_n\cdot\nabla)^2 \vec u.
\end{equation}
The last term in \eqref{eq:D2} is the trigonal warping term. This trigonal warping terms needs to be included when we compute the Christoffel symbols (as well as spin connection) and the curvature, since we work to second order in the derivatives of the strain field $u_i(\mx)$. This subtlety was missed in the previous work \cite{Juan:2012}. We then have 
\begin{equation}
	\frac{1}{a^2}\bigg(\vec u(\vec x+\vec l_n)-\vec u(\vec x)\bigg)\cdot \mathbf{l}_n\approx \frac{l_n^il_n^j u_{ij}}{a^2}+\frac{l_n^il_n^jl_n^k\partial_i u_{jk}}{2a^2}
\end{equation}
where we have defined
\begin{equation}
	u_{ij}\equiv \partial_i u_j
\end{equation}
It will be useful to define the following matrices
\begin{equation}
	\frac{1}{a}\sum_{n=1}^3 l_n^i=0,
\end{equation}
\begin{equation}
	\frac{1}{a^2}\sum_{n=1}^3 l_n^i l_n^j=\frac{3}{2}\delta^{ij},
\end{equation}
\begin{equation}
	\frac{1}{a^3}\sum_{n=1}^3 l_n^i l_n^jl_n^k=-\frac{3}{4}K^{ijk},
\end{equation}
\begin{equation}
	\frac{1}{a^4}\sum_{n=1}^3 l_n^i l_n^jl_n^kl_n^l=\frac{3}{8}L^{ijkl},
\end{equation}
The matrices $K$ and $L$ are completely symmetric in all their indices and hence have 4 and 5 independent components respectively. All the entries of these matrices are integers. The independent entries are
\begin{equation}
	K^{111}=0, \ K^{112}=-1,\ K^{122}=0,\ K^{222}=1
\end{equation}
and
\begin{align}
	&L^{1111}=3, \ L^{1112}=0, \ L^{1122}=1, \ L^{1222}=0,\\ &\ L^{2222}=3\nonumber
\end{align}
We now want to calculate the spatially-dependent Fermi velocity and the gauge field to second order. Recall the expressions
\begin{align}
	\label{eq:spatial_vF_lin}
	\tilde{v}^{ij}(\x)=\sum_{n}\frac{2}{3t_0 a^2}l_n^il_n^j\left(t_n(\x)-\frac{1}{2}l^k_n \d_k t_n(\x)\right),
\end{align}
\begin{equation}
	\label{eq:A_expr_lin}
	A^{s}_i(\x)=\sum_{n}\frac{2}{3 t_0 a^2}\epsilon_{ij}l^j_n\left(t_n(\x)-\frac{1}{2}l^k_n \d_k t_n(\x)\right)
\end{equation}
Plugging in, we find 
\begin{equation}
	\tilde{v}^{ij}(\x)=\delta^{ij}-\frac{\beta}{4}L^{ijkl}u_{kl}
	\label{eq:App_vF_lin}
\end{equation}
\begin{equation}
	\tilde v^{ij}=\delta^{ij}-\frac{\beta}{4}(u^{ij}+u^{ji}+\delta^{ij}u^{kk})
	\label{eq:v_1st_lin}
\end{equation}
Furthermore, we find
\begin{equation}
	A^{s}_i(\x)=\frac{\epsilon_{ij}}{a}\frac{\beta}{2}K^{jkl}u_{kl}
	\label{eq:App_gauge_field_lin}
\end{equation}
\begin{equation}
	\vec A^{s}=\frac{\beta}{2a}\begin{pmatrix}
		u_{yy}-u_{xx}\\
		u_{xy}+u_{yx}
	\end{pmatrix}.
	\label{eq:A_1st_lin}
\end{equation}
\eqref{eq:v_1st_lin} and \eqref{eq:A_1st_lin} agree with the results in \cite{Juan:2012}. However, the fact that the results coincide is non-trivial. We note that compared to \cite{Juan:2012} we have included the trigonal warping term in \eqref{eq:D2}. We also have the additional derivative terms in \eqref{eq:vb} and \eqref{eq:At}. These additional derivative terms exactly cancel off the contribution from the trigonal warping. Moreover, the combination of the trigonal warping and the extra terms in \eqref{eq:vb} and \eqref{eq:At} also eliminate the extra gauge potential $\tilde{\mathbf{A}}_a$ of Ref.~\cite{Zubkov:2015}. The $\tilde{\mathbf{A}}_a$  term of Ref.~\cite{Zubkov:2015}, with the presence of $\epsilon_{ab}$, is forbidden by the reflection symmetry, see  \cite{First_order} and \cite{Winkler:2010} for more detailed arguments.

\section{Second-order calculation: exponential tunneling}
\label{sec:App_2nd_order}
We assume that each lattice site is displaced as
\begin{equation}
	\vec{x}\rightarrow\vec{x}+\vec{u}({\vec{x}})
\end{equation}
Two lattice sites that were previously separated by $\vec l_n$ will now be separated by
\begin{equation}
	\vec l_n^\prime=\vec l_n+\vec{u}(\vec x+\vec l_n)-\vec u(\vec x)\approx \vec l_n +(\vec l_n\cdot\nabla) \vec u+\frac{1}{2}(\vec l_n\cdot\nabla)^2 \vec u
\end{equation}
We now note that there are two small parameters in the problem. Defining $a\equiv |\vec l_n|$ and $L$ as the total size of the system,
\begin{equation}
	\frac{(\vec l_n\cdot \nabla )\vec u}{a}\sim \frac{\vec u}{L}
\end{equation}
\begin{equation}
	\bigg(\frac{(\vec l_n\cdot \nabla )\vec u}{a}\bigg)\sim\bigg( \frac{\vec u}{L}\bigg)^2
\end{equation}
\begin{equation}
	\frac{(\vec l_n\cdot\nabla)^2 \vec u}{a}\sim \frac{\vec u}{L}\frac{a}{L}
\end{equation}
We assume that the two small parameters $\vec u/L$ and $a/L$ are of a similar order of smallness and expand to second-order in these small parameters. Expanding, we find after some algebra
\begin{equation}
	\frac{|\vec l_n^\prime|-a}{a} \approx \frac{l_n^il_n^j u_{ij}}{a^2}+\frac{l_n^il_n^j u_{ik}u_{jk}}{2a^2}-\frac{l_n^il_n^jl_n^kl_n^mu_{ij}u_{km}}{2a^4}+\frac{l_n^il_n^jl_n^k\partial_i u_{jk}}{2a^2}
\end{equation}
where we have defined
\begin{equation}
	u_{ij}\equiv \partial_i u_j
\end{equation}
To second order
\begin{equation}
	\bigg(\frac{|\vec l_n^\prime|-a}{a}\bigg)^2\approx  \frac{l_n^il_n^j l_n^kl_n^mu_{ij}u_{km}}{a^4}
\end{equation}
The hopping will depend on the distance between the sites.
\begin{equation}
	t_n(\vec x)\approx t_0\cdot \bigg[1-\beta \frac{|\vec l_n^\prime|-a}{a} +\kappa \bigg(\frac{|\vec l_n^\prime|-a}{a}\bigg)^2\bigg]
\end{equation}
where the coefficients $\beta$ and $\kappa$ are defind by
\begin{equation}
	\beta=-\frac{a}{t_0}\frac{\partial t}{\partial r}\bigg\rvert_{r=a}, \qquad \kappa=\frac{a^2}{2t_0}\frac{\partial^2 t}{\partial r^2}\bigg\rvert_{r=a}
\end{equation}
For the typical exponentially decaying hooping, which is a good approximation in both the case of real graphene and its photonic analogue, we have $t=t_0e^{-\beta(r-a)/a}$ and hence $\kappa=\beta^2/2$. In addition to the matrices defined in the previous section, it will be useful to define the following matrices
\begin{equation}
	\frac{1}{a^5}\sum_{n=1}^3 l_n^i l_n^jl_n^kl_n^ll_n^m=-\frac{3}{16}M^{ijklm},
\end{equation}
\begin{equation}
	\frac{1}{a^6}\sum_{n=1}^3 l_n^i l_n^jl_n^kl_n^ll_n^ml_n^o=\frac{3}{32}N^{ijklmo}.
\end{equation}
The matrices $M$ and $N$ are completely symmetric in all their indices and hence have 6 and 7 independent components respectively. All the entries of these matrices are integers. The independent entries are
\begin{align}
	&M^{11111}=0, \quad M^{11112}=-3, \quad M^{11122}=0, \\ &M^{11222}=-1, \quad M^{12222}=0, 
	M^{22222}=5 \nonumber
\end{align}
and
\begin{align}
	&N^{111111}=9, \quad N^{111112}=0, \quad N^{111122}=3,\\ &N^{111222}=0, \quad N^{112222}=1,\nonumber
	N^{122222}=0,\\ &N^{222222}=11\nonumber.
\end{align}
We now want to calculate the spatially-dependent Fermi velocity and the gauge field to second order. Recall the expressions
\begin{align}
	\label{eq:spatial_vF}
	\tilde{v}^{ij}(\x)=\sum_{n}\frac{2}{3t_0 a^2}l_n^il_n^j\left(t_n(\x)-\frac{1}{2}l^k_n \d_k t_n(\x)\right),
\end{align}
\begin{equation}
	\label{eq:A_expr}
	A^{s}_i(\x)=\sum_{n}\frac{2}{3 t_0 a^2}\epsilon_{ij}l^j_n\left(t_n(\x)-\frac{1}{2}l^k_n \d_k t_n(\x)\right)
\end{equation}
Plugging in, we find	
\begin{equation}
	\tilde{v}^{ij}(\x)=\delta^{ij}-\beta\bigg(\frac{1}{4}L^{ijkl}u_{kl}+\frac{1}{8}L^{ijkl}u_{km}u_{lm}-\frac{1}{32}N^{ijklmo}u_{kl}u_{mo}\bigg)+\frac{\kappa}{16}N^{ijklmo}u_{kl}u_{mo}
	\label{eq:App_vF}
\end{equation}
To first order in strain,
\begin{equation}
	\tilde v^{ij}=\delta^{ij}-\frac{\beta}{4}(u^{ij}+u^{ji}+\delta^{ij}u^{kk})
	\label{eq:v_1st}
\end{equation}
Furthermore, we find
\begin{equation}
	A^{s}_i(\x)=\frac{\epsilon_{ij}}{a}\bigg[-\beta\bigg(-\frac{1}{2}K^{jkl}u_{kl}-\frac{1}{4}K^{jkl}u_{km}u_{lm}+\frac{1}{16}M^{jklmo}u_{kl}u_{mo}\bigg)-\frac{\kappa}{8}M^{jklmo}u_{kl}u_{mo}\bigg]
	\label{eq:App_gauge_field}
\end{equation}
Again, if we work to first order in strain, we find
\begin{equation}
	\vec A^{s}=\frac{\beta}{2a}\begin{pmatrix}
		u_{yy}-u_{xx}\\
		u_{xy}+u_{yx}
	\end{pmatrix}.
	\label{eq:A_1st}
\end{equation}
To this order, \eqref{eq:v_1st} and \eqref{eq:A_1st} agree with the results in \cite{Juan:2012}. However, we have calculated the higher-order corrections. For the strain profile \eqref{eq:strain_profile} and setting $\kappa=\beta^2/2$ we find curvature
\begin{equation}
	\cK=-\frac{4}{a^2}\bigg(\beta\frac{au_B}{L}\bigg)^2
	\label{eq:K_exp}
\end{equation}
and magnetic field	
\begin{equation}
	B=\frac{4}{a^2}\beta\frac{au_B}{L}-\frac{4}{a^2}\bigg(\beta\frac{au_B}{L}\bigg)^2.
	\label{eq:B_exp}
\end{equation}

\bibliography{SG}

\end{document}